\documentclass[aps,prl,reprint,amsmath,amssymb,superscriptaddress]{revtex4-1}
\usepackage{color}
\usepackage{graphicx}
\usepackage{subfigure}
\usepackage{comment}
\usepackage{hyperref}
\usepackage{braket}
\usepackage[normalem]{ulem}

\graphicspath{{./figures/}}

\definecolor{gold}{rgb}{0.83, 0.69, 0.22}

\newcommand{\ssec}[1]{\emph{#1}.---}
\definecolor{mygreen}{RGB}{61,145,64}

\definecolor{violet}{RGB}{100,0,200}

\begin{document}

%\begin{CJK*}{UTF8}{}
\title{Stable solid molecular hydrogen above 900K from a machine-learned potential trained with diffusion Quantum Monte Carlo}
\author{Hongwei Niu}
\affiliation{Department of Astronautical Science and Mechanics, Harbin Institute of Technology, Harbin, Heilongjiang 150001, China}
\author{Yubo Yang}
\affiliation{Center for Computational Quantum Physics, Flatiron Institute, New York, New York 10010, USA}
\affiliation{Department of Physics, University of Illinois, Urbana, Illinois 61801, USA}
\author{Scott Jensen}
\affiliation{Department of Physics, University of Illinois, Urbana, Illinois 61801, USA}
\author{Markus Holzmann}
\affiliation{Univ. Grenoble Alpes, CNRS, LPMMC, 38000 Grenoble, France}
\author{Carlo Pierleoni}
\affiliation{Department of Physical and Chemical Sciences, University of L'Aquila, Via Vetoio 10, I-67010 L'Aquila, Italy}
\author{David M. Ceperley}
\affiliation{Department of Physics, University of Illinois, Urbana, Illinois 61801, USA}
\date{\today}
\begin{abstract}
We survey the phase diagram of high-pressure molecular hydrogen with path integral molecular dynamics using a machine-learned interatomic potential trained with Quantum Monte Carlo forces and energies. Besides the HCP and C2/c-24 phases, we find two new stable phases both with molecular centers in the Fmmm-4 structure, separated by a molecular orientation transition with temperature. The high temperature isotropic Fmmm-4 phase has a reentrant melting line with a maximum at higher temperature (1450K at 150GPa) than previously estimated and crosses the liquid-liquid transition line around 1200K and 200GPa.
\end{abstract}
\pacs{}
\maketitle
%\end{CJK*}

%\section{Introduction} \label{sec:melth2-intro}
\ssec{Introduction}
% what controversies are there in experiment?
Experimental methods for probing the phase diagram of high-pressure hydrogen are limited. At room temperature and below, the diamond anvil cell (DAC) has allowed exploration for pressures up to roughly  450 GPA~\cite{RangaI.F.2017,Eremets2019,Loubeyre2020}. However, the small size of the cell and fragility of the sample limit experimental probes to low-power optics such as infrared and Raman spectroscopy~\cite{Loubeyre2022}. %\paul{reference change left and right} 
Hydrogen weakly scatters X-Rays~\cite{Ji2019}, making structural determination difficult. Recently, direct measurement of the structure of solid molecular hydrogen has been achieved up to pressures of 254 GPa~\cite{Akahama2010,Ji2019}. 
At temperatures below 100K, pressures above 400GPa can be achieved in DAC, but X-ray structural determination is not yet available.
At higher temperatures, shock wave compression methods have achieved higher pressures, but due to the transient nature of these experiments, acquiring and analyzing shock-wave data is challenging. Notably, one cannot directly measure temperature, which may cause difficulty interpreting results~\cite{Celliers2018,Knudson2004,Knudson2017}.
Given the experimental difficulties, accurate simulations are necessary to inform and complement experimental works~\cite{Pierleoni2016b}.

% what uncertainty has recently emerged in the the phase diagram of hydrogen?
% what controversies are there in calculations?
Simulation of high-pressure hydrogen requires accurate methods both in the description of the electronic ground-state Born-Oppenheimer (BO) potential energy surface (PES) and the inclusion of nuclear quantum effects beyond the harmonic approximation. %Current approaches require some compromise between accuracy and practicality. %Without experimental structural information, m
Many calculations have been performed on structures found with density functional theory (DFT) based random structure searches~\cite{Pickard2007}. %Constrained by the computational cost, these searches are often performed with classical protons missing saddle-point structures that can be stabilized by nuclear quantum effects~\cite{Monserrat2016}.
The Perdew-Burke-Ernzerhof (PBE) exchange-correlation functional often used in DFT studies incorrectly predicts that the molecular structures in phase III are metallic~\cite{Drummond2015}. %However, its use in conjunction with classical molecular dynamics (MD) has produced reasonable results for the LLT due to error cancellation~\cite{Morales2013a}. %
Benchmarks using diffusion Monte Carlo (DMC)~\cite{Clay2016} have established that the vdW-DF1 functional is the best compromise for molecular hydrogen at pressures above 100GPa.
But exploring large pressure and temperature ranges using first-principles methods is so time consuming that establishing convergence with respect to supercell size and trajectory length is difficult to achieve. 
%\CP{\it{the following sentence concerns a completely different problem that we do not address in the paper. I think we can mention it but here it just cut the flow of narrative.}} Further, some experimentally measurable properties, such as emission, absorption, infrared and Raman spectra, heat conductivity, and diffusion constants are difficult to calculate in the most accurate electronic methods, e.g. diffusion quantum Monte Carlo.

Calculations using the most accurate simulation technique, Coupled Electron Ion Monte Carlo (CEIMC), which calculates electronic energies using explicitly correlated wave functions, find that solid molecular hydrogen remains stable at higher temperatures than indicated by simulations based on PBE forces and higher than experimental estimates. %\david{MORE?? See the discussion and Fig. () in the SM.  } 
However, the number of atoms and the length of trajectories accessible by CEIMC is limited by computer resources. This is particularly worrying for disoriented molecular phases: alternative algorithms for studying the melting line are needed.

In recent years, machine-learned (ML) interatomic potentials have emerged as a promising tool providing a balance between accuracy and efficiency thus allowing for accurate but less expensive calculations~\cite{Han2018,Zhang2018a,Zhang2018} that can address the temporal and spatial limitations of first-principle simulations. ML methods have recently been applied to dense hydrogen.
However, there are several conflicting theoretical results regarding dissociation, melting, and the critical point obtained with the various simulation methods
~\cite{Zong2020,Cheng2020,karasiev2021,Tirelli2022}.

% how can we help?
In this study, we perform large-scale simulations of molecular hydrogen using quantum protons with a ML approximation to the DMC-BO-PES for pressures from 50 to 220 GPa.

%\section{Method} \label{sec:melth2-method}
\ssec{Training data} 
%To investigate the phase diagram of molecular hydrogen using a machine-learned potential with DMC accuracy, 
Our training data consists of about $10^5$ hydrogen configurations, each with 96 protons, obtained from classical molecular dynamics (MD), path integral molecular dynamics (PIMD) and CEIMC simulations, the majority of which are of molecular hydrogen. Their pressures and temperatures are shown in the supplementary material (SM), which also include Refs.~\cite{RMP2012,Giannozzi2009,Enkovaara2017,Kim2018,Kent2020,Chiesa2005,Chiesa2006,Holzmann2016,Bartok2010,Pachucki2008,Gorelov2020,LAMMPS,kapil2019pi}.
The total energy, atomic forces, and cell stresses are evaluated with DFT using both PBE and vdW-DF1 functionals. For a subset of approximately 20,000 configurations the energy and forces are estimated by DMC using the CCZ estimator \cite{Chiesa2005} to a  statistical error bar of 0.1 meV and  130 meV/\AA ~ respectively. %(fixed-node DMC calculations with an optimized Slater-Jastrow wavefunction. 
As we show in the SM, the DMC configurations are over-complete so that the statistical error in the forces of the ML models is about about 80meV/\AA. This data is publicly available \cite{ythub}. 

\ssec{Hierarchical $\Delta$-learning}
The potential energy model is a sum of three terms:
\begin{equation} \label{eq:final-model}
E = E_{\text{pair}} + \Delta E_{DFT} + \Delta E_{DMC}
\end{equation}
where $E_{\text{pair}}$ is the contribution from a proton pair potential using only the repulsive part of the exact BO energy of a H$_2$ molecule~\cite{Kolos1964}. 
This allows an accurate description of the internal motion of a single H$_2$
%The pair potential imposes the correct short-distance asymptotic behavior of the two-body interaction 
without having to learn it from the training data.

The next two terms in eq. (\ref{eq:final-model}) are determined by a ML model.
$\Delta E_{DFT}$ is a fit to DFT energy, forces, and stresses minus the pair potential contribution over all of the configurations. 
This fit should capture the overall band structure and electron correlation effects as described by the density functional.
Finally, $\Delta E_{DMC}$ is determined by fitting $E - E_{\text{pair}} - \Delta E_{DFT}$ to the subset of configurations that have QMC forces.  The difference between the DFT energy and the QMC energy is expected to be small and a smooth function of positions and thus requires a simpler network. 
Final results of the DMC trained model are fairly independent of the DFT functional used whether PBE or vdW-DF1. The results which we present here used vdW-DF1 functional.
This procedure makes optimal use of the computationally expensive and noisy QMC forces.
%This captures the remaining effects of correlation  to produce our final hierarchical DMC model \david{what is the difference between the 2 DEEPMD models?}

\begin{figure}[tbh]
\includegraphics[scale=0.28]{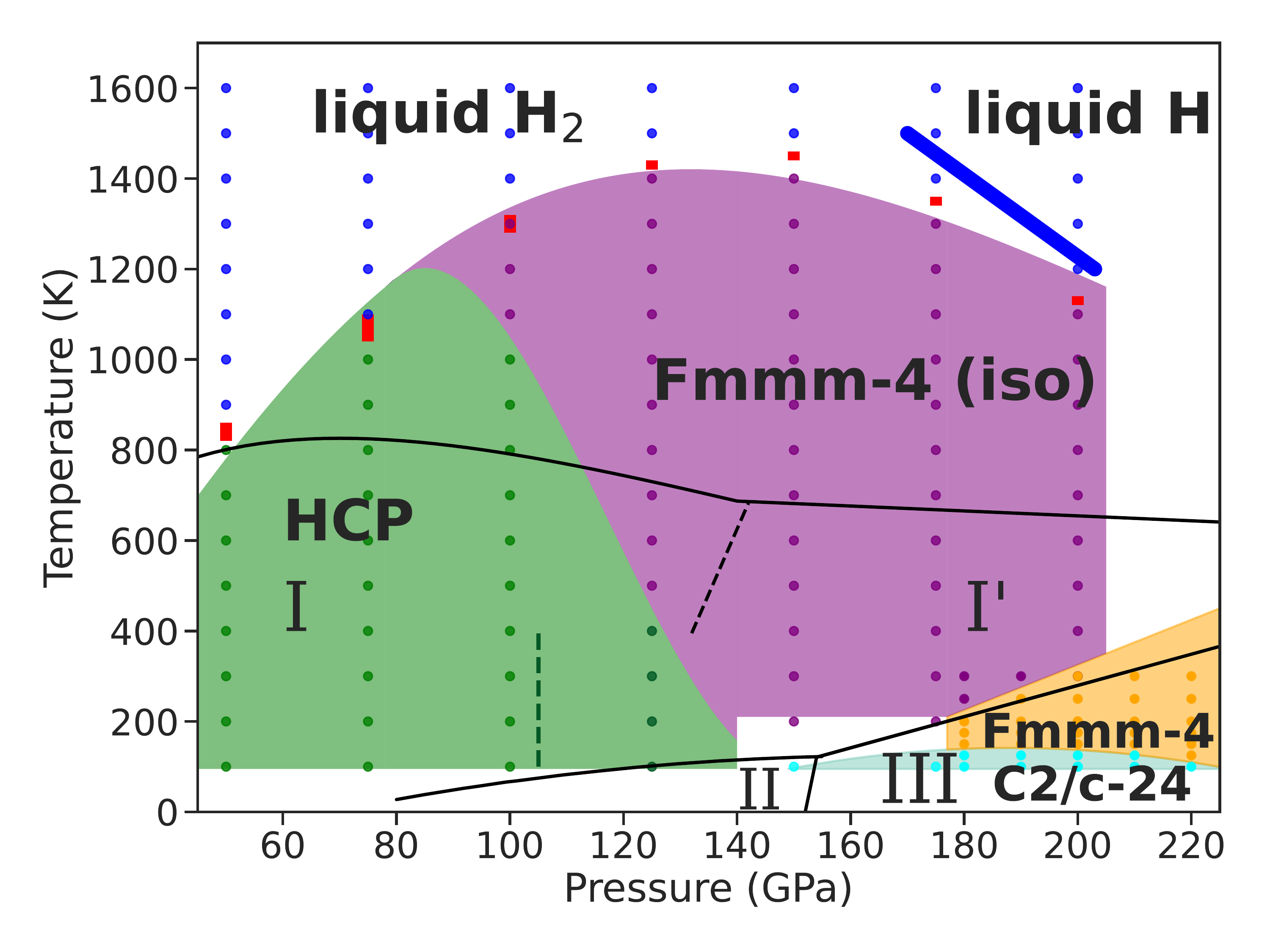}
\caption{Phase diagram of dense hydrogen.   The dots indicate the (P,T) values where we ran PIMD simulations using a DMC-trained DPMD interatomic potential.
Colors indicate the identified phase: dark blue (liquid), green (HCP), purple (Fmmm-4 isotropic), orange (Fmmm-4 oriented) and cyan (C2/c-24). The green dashed line at 105 GPa indicates a crossover within HCP to an in-plane orientation.  
The thick blue line is the estimate of the transition from liquid  H$_2$ to liquid H from ref. \cite{Pierleoni2016b}.
Red bars are estimates of the melting temperature from two-phase coexistence simulation.
The black lines are experimental estimates of phase boundaries, solid lines for the melting, dashed line for I-I' transition ~\cite{Zha2017}.
}
\label{fig:phase-diagram}
\end{figure}

To construct the two ML models in Eq. \ref{eq:final-model}, we use the Deep Potential Molecular Dynamics (DPMD)~\cite{Han2018,Zhang2018a,Zhang2018} framework: the energy is a sum of atomic contributions, similar to the embedded atom method (EAM)~\cite{Daw1983,Daw1984}.
The contribution from one atom is calculated from its local atomic environment, represented by a real-valued vector that is invariant with respect to translation, rotation, and permutation. See the SM for further model and hyperparameter details.
% details of electronic structure calculations and molecular dynamics

%\section{Results} \label{sec:melth2-results}
%\subsection{Phase Diagram} 

\ssec{PIMD Simulation}
The nuclear quantum effects of protons are substantial, thus PIMD simulations using a NPT ensemble are used to explore the phase diagram of molecular hydrogen with the DMC-trained potential. 
The melting line shown in Fig.~\ref{fig:phase-diagram}
was obtained using a two-phase method~\cite{BELONOSHKO1994}: the PIMD simulations started from a half-liquid half-solid configuration consisting of 3072 protons. See the SM for details of the analysis and comparison between MD and PIMD results.

\ssec{Equation of State}
\begin{figure}
\centering
\includegraphics[width=\linewidth]{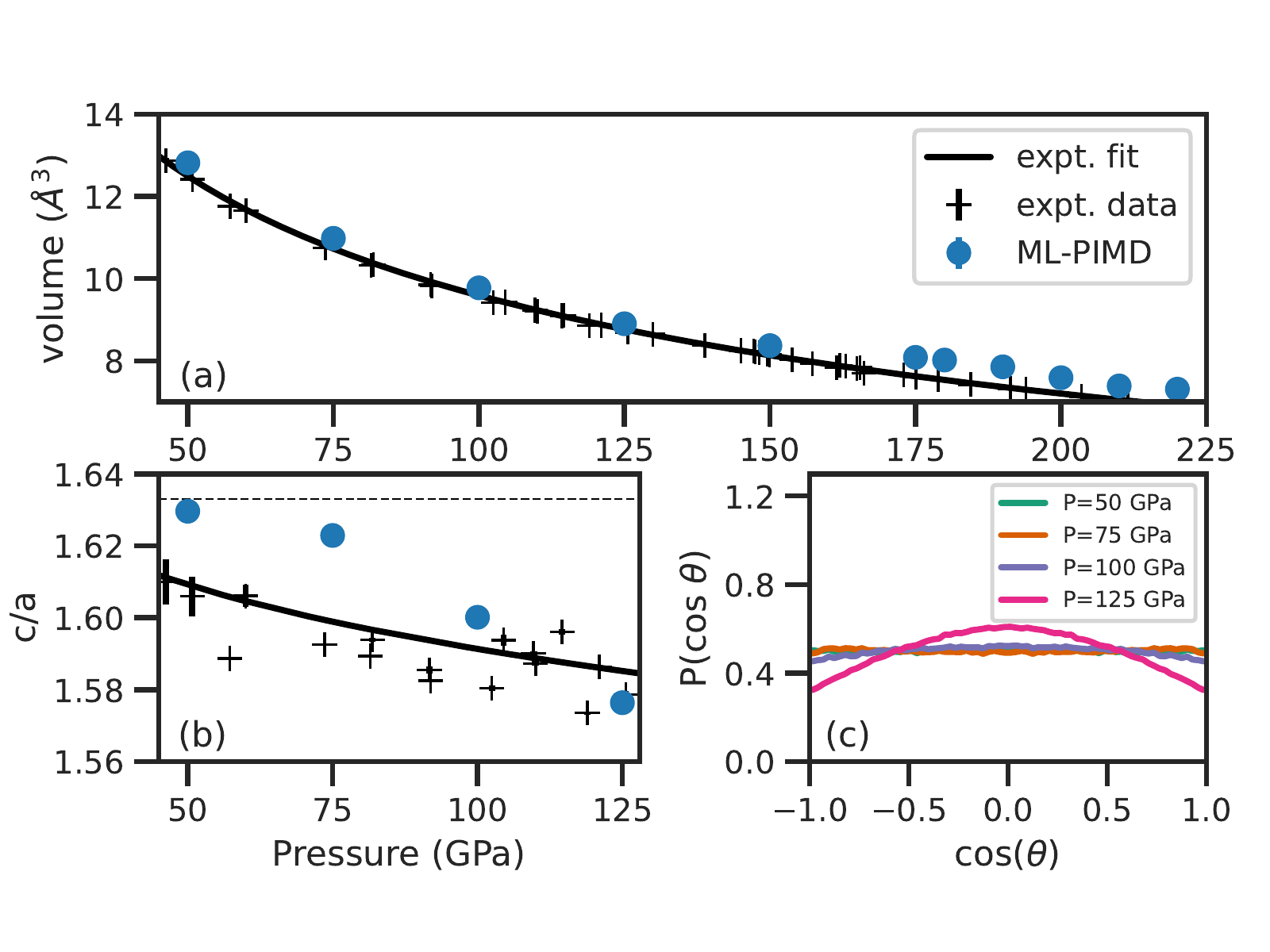}
\caption{(a) Unit cell volume and (b) $c/a$ ratio, vs. pressure in GPa both at 300K. Blue dots are the results of PIMD simulation. The crosses are experimental XRD measurements Ref.~\cite{Jinnouchi2019} and the lines are fits to experimental data. (c) The distribution of the molecular angle with the basal plane at various pressures within HCP at 300K from PIMD.}
\label{fig:ca}
\end{figure}  
Comparison of our results with X-ray scattering experiments of molecular hydrogen up to 225GPa at room temperature~\cite{Ji2019} is shown in Fig.~\ref{fig:ca}. The overall comparison of the equation of state is excellent. Experiment identified the structure as hcp, but we find the hexagonal symmetry is broken above 150GPa  as discussed next. The c/a ratio decreases with increasing pressure in both simulation and experiment. We find an anisotropy developing in the molecular orientation as  the c/a ratio deviates from the closed packed limit.

\ssec{Solid structures}
To identify the stable crystal structures in the PIMD trajectories  we use several techniques: the molecular center of mass structure factor, the protonic structure factor and the molecular bond orientational distribution. Snapshots are also quenched to determine the ideal crystal structure and then the crystal symmetries are found using the FINDSYM package ~\cite{FINDSYM}.  %from the other candidate structures (HCP and C2/c-24),

\begin{figure*} [tbh]
\hspace*{\fill}%
\setcounter {subfigure} {0} (a){
\includegraphics[width=0.42\linewidth]{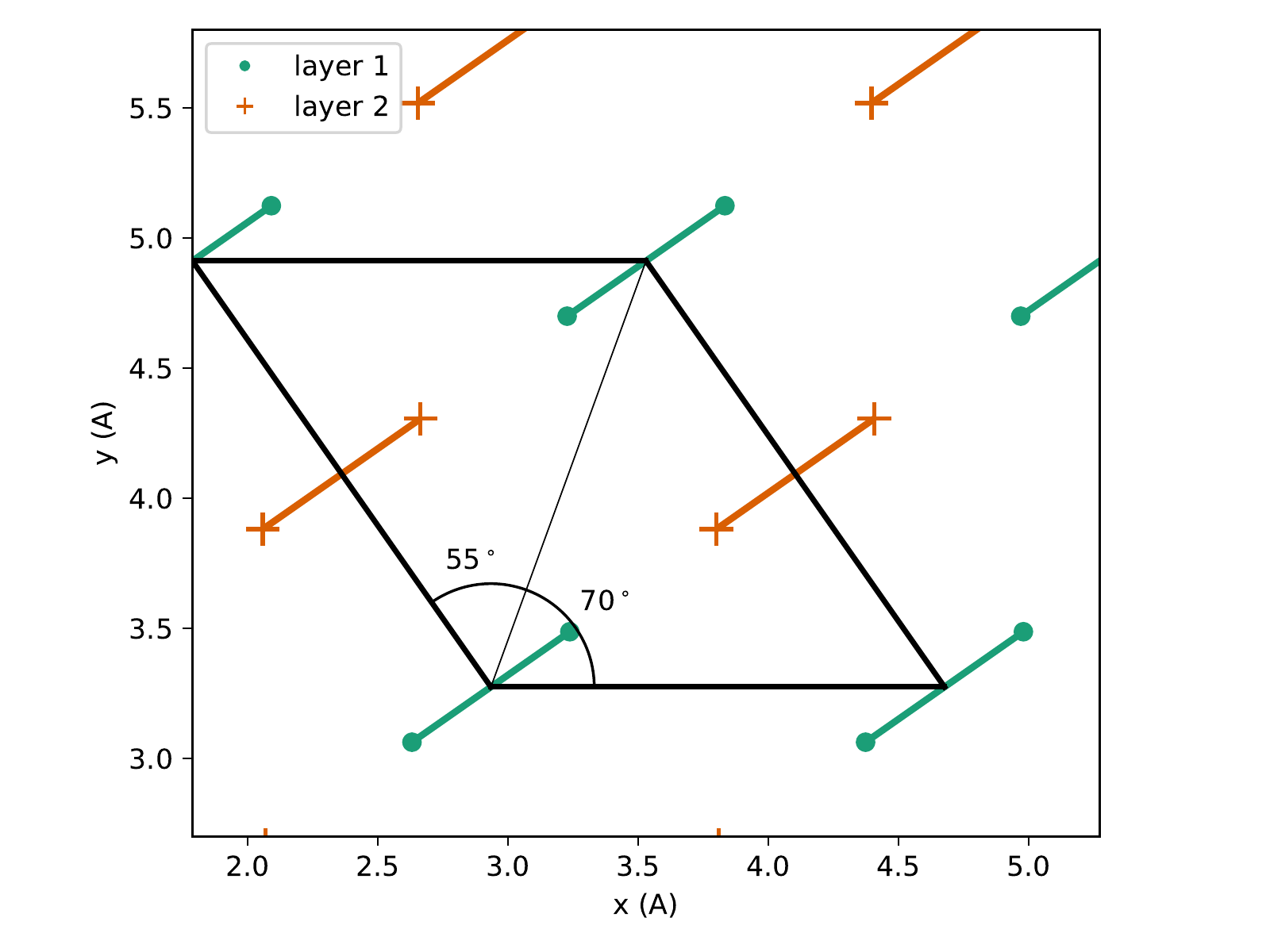}
}
\setcounter {subfigure} {0} (b){
\includegraphics[width=0.42\linewidth]{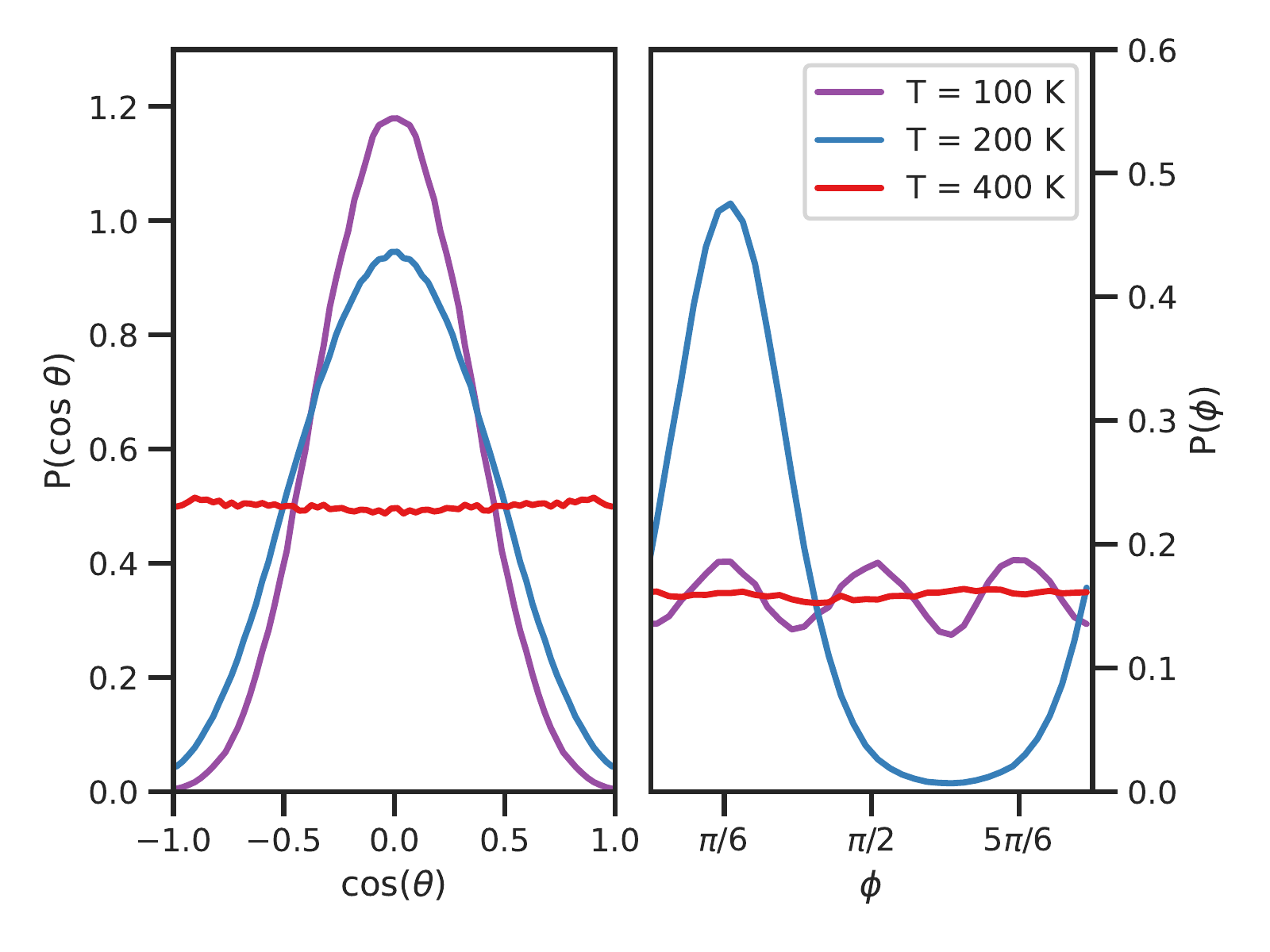}
}
\hspace*{\fill}%
\caption{
(a) The oriented Fmmm-4 structure with layer 1 (green) and layer 2 (orange).
The primitive cell (shown in black) contains one molecule per layer all lying in the basal plane.
Layer 2 is shifted by half a primitive lattice vector from the HCP site. Also the hexagonal symmetry within the basal plane is broken.
(b) Histograms of molecular orientation at P=$200$ GPa where $\theta$ is the angle with respect to the basal plane and $\phi$ is the azimuthal angle.
There are three distinct in-plane orientations in the C2/c-24 structure at $100$ K, one orientation in the Fmmm-4 structure at $200$ K.
Orientation order is absent in the isotropic Fmmm-4 phase at $400$ K and above. 
}
\label{fig:solid_structure}
\end{figure*}

As shown in Fig.~\ref{fig:solid_structure} we find the dominant structure above 400K and 120GPa to be the isotropic Fmmm-4 with 2 molecules per unit cell without preferential orientation.
In the HCP structure, a molecular center of the second layer lies above the center of the equilateral triangle formed by three molecular centers in the first layer, while
in Fmmm-4, the molecular center of the second layer lies above the edge of that triangle. The hexagonal symmetry of the basal plane is broken so that the angle between the two primitive vectors becomes 70.5$^o \pm 0.2^o$ throughout this phase. For T $\leq$ 400K the molecules become oriented in the basal plane as shown in  Fig.~\ref{fig:solid_structure}. 
For T $<$ 200K we find the structure to be C2/c-24, one of the candidate structures predicted in \cite{Pickard2007}. This is again an HCP lattice of molecular centers with oriented molecules; there are 4 distinct layers with a total of 12 molecules per unit cell.

\ssec{Phase diagram} 
Figure~\ref{fig:phase-diagram} shows the  phase diagram of dense hydrogen using the DMC-trained ML model compared to experimental results of Ref.~\cite{Zha2017} 
based on Raman measurements.
At 100 K, we find the HCP phase to be stable up to 150 GPa. Below 100GPa, the molecules are isotropic, but as the pressure increases their orientation begins to favor the basal plane as shown in \ref{fig:ca}(c). The azimuthal orientation remains isotropic.

We find a solid transition to an orientationally ordered C2/c-24 structure at around 150GPa.
 %the Raman and IR signatures of the 
This change has been seen in a recent experiment~\cite{Loubeyre2022}. 
The C2/c-24 structure is absent from our training set, so its spontaneous emergence demonstrates the predictive power of the model.
A previous ground-state QMC of a perfect crystal puts the transition to the C2/c-24 at 250 GPa~\cite{Drummond2015} while for 150GPa $\leq$ P $\leq$ 250GPa  P2$_1$/c-24 is the ground-state.
We attribute this difference to treatment of the nuclear quantum effect: we observe large-amplitude librations of the H$_2$ molecules (see Fig.~\ref{fig:solid_structure}(b)) which are inadequately treated within the vibrational self-consistent field method underlying the prediction of Ref.~\cite{Drummond2015}.

We find that for $T>150$ K and $P>160$ GPa, the C2/c-24 crystal transitions into an oriented Fmmm-4 structure. Even in this phase, the molecules have large librations as shown in Fig.~\ref{fig:solid_structure}(b). 
At even higher temperature, under conditions similar to the experimental phase I'-III boundary, the orientational order is lost in favor of an isotropic Fmmm-4 phase (see Fig.~\ref{fig:solid_structure}). We see a sharp orientational transition in temperature and observe hysteresis, signaling the first-order nature of the transition.

Previous structure searching \cite{Pickard2007} did not favor the Fmmm-4 phases  because they are stable only above 200K, the molecules are highly disoriented, so that estimates of the zero point energy based on a perturbation expansion about a perfect crystal is difficult, and density functionals may not be precise enough to distinguish between the hydrogen structures. Note that in this region of temperature and pressure (200K and 150GPa), \emph{five } different crystal phases are relevant. Further studies will be needed to predict their stability and understand the experimental observations.

\ssec{Molecular melting curve} The ML model finds the molecular HCP solid structure melts at 800 K at 50GPa,  consistent with experiment~\cite{Zha2017} and the melting temperature increases to about 1100K at about 90GPa. Above this pressure, we find that the HCP structure transitions to the isotropic Fmmm-4 phase before melting,
with a  maximum melting temperature 1450 K at 150 GPa. Our result suggests the presence of a triple point HCP-Fmmm-4(iso)-Liquid at around 90GPa and 1100K and a HCP-Fmmm-4(iso) transition line declining rapidly with pressure. Our prediction of melting is in striking disagreement with experiments.
The experimental estimates of melting for pressures above 100GPa are indirect: they do not show absence of long-range density order or shear stress, but rather rely on a change in the Raman spectra ~\cite{Subramanian2011,Howie2015,Zha2017}, in the rate of heating of the sample~\cite{Deemyad2008}, or of its dielectric properties~\cite{Eremets2009}.
The increase in melting temperature that we find can be attributed to the higher stability of the isotropic Fmmm-4 structure with respect to the liquid.

\begin{figure}[h]
\includegraphics[width=\linewidth]{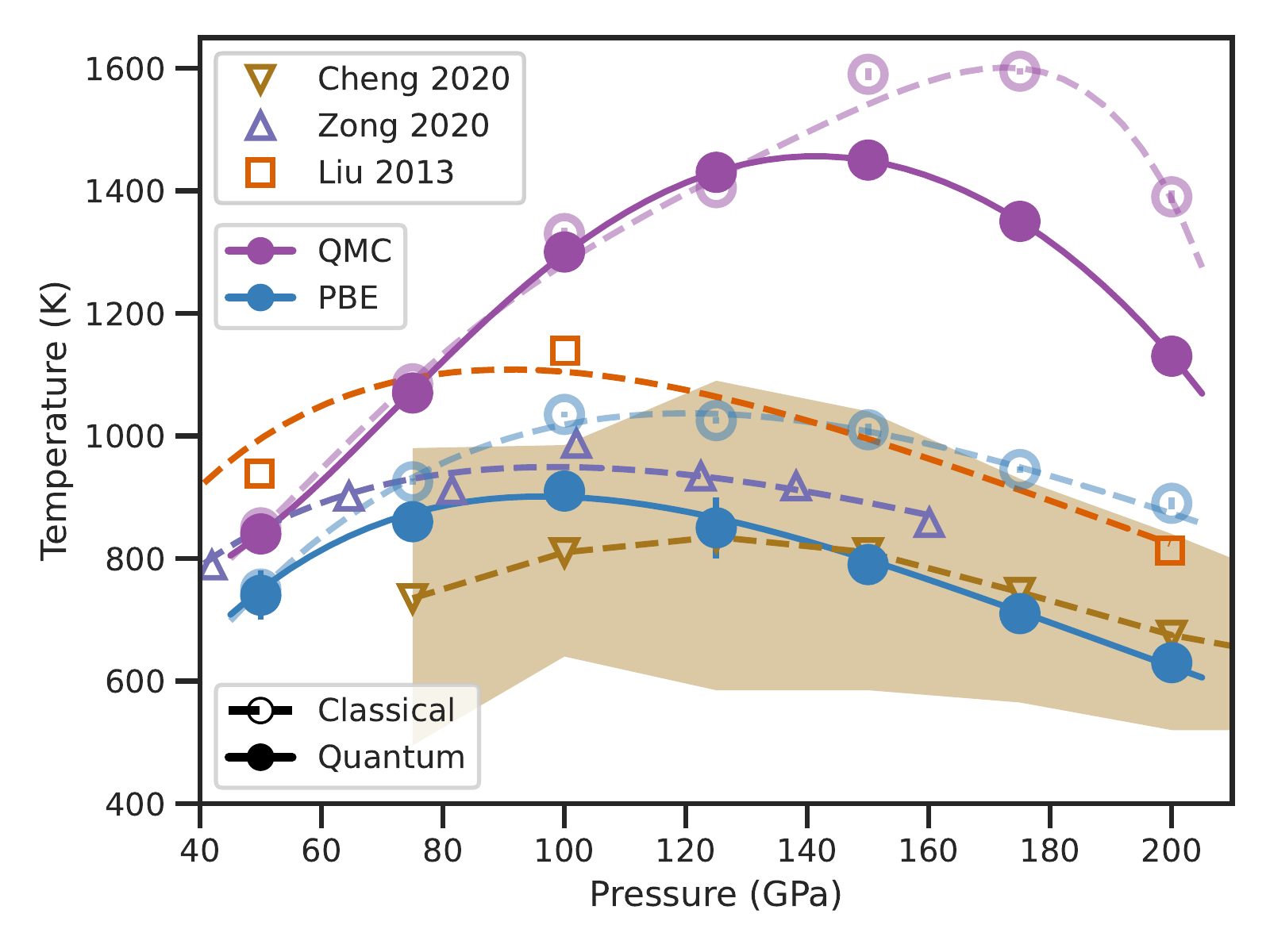}
\caption{
Comparison of melting lines from various calculations. Circles are our two-phase MD results using ML constrained by data from either PBE forces (blue) or QMC forces (purple). Open symbols represent classical protons, closed symbols quantum protons.
Previous PBE-DFT predictions, all using classical protons, are reported: gold downward triangles ref.~\cite{Cheng2020}, light purple upward triangles ref.~\cite{Zong2020}, orange squares ref.~\cite{Liu2013}). Lines (either continuous or dashed) are guides for the eyes. The shaded region is the uncertainty estimated in Cheng \textit{et al.}~\cite{Cheng2020}.
}
\label{fig:calc-melt}
\end{figure}

Figure~\ref{fig:calc-melt} shows the predictions of the melting line from DMC-trained and PBE-trained models using both classical and quantum proton simulations. 
Above 100GPa the melting temperature is very sensitive to the underlying electronic PES or the mass of the protons. Closed symbols represent quantum protons, while open symbols classical protons. 
Figure~\ref{fig:calc-melt} also reports previous melting calculations based on the PBE functional, either by DFT-MD\cite{Liu2013} or by DFT-trained ML potential\cite{Zong2020,Cheng2020}.
Our ML predictions of classical protons based on the PBE functional (blue open circles) are in reasonable agreement with previous studies giving us confidence in the reproducibility of the computational procedure.
We observe a divergence of the classical and quantum melting curves around 120 GPa consistent with the DFT calculations of the D$_2$ and H$_2$ melting curves by Caillabet et al.~\cite{Caillabet2011}.
%The good agreement between our ``classical'' PBE melting temperatures from 100 to 200 GPa with those obtained by DFT-MD on a rather large cell (brown line) is especially reassuring as we both used the two-phase coexistence method in large simulation cells.
%When compared to other machine learning results, our ``classical'' PBE melting temperatures are consistent with the orange line and the upper bound of blue region.

The melting maximum is at about 150GPa for the DMC-trained model and at about 100GPa for the PBE-trained model with either quantum or classical protons. Melting based on vdW-DF1 trained model (see SM) lies in-between the PBE and DMC models. The nuclear quantum motion lowers the melting temperature by roughly 100-200 K above 100 GPa. These large differences highlight the importance of accurately treating both the electronic structure and nuclear quantum effect in dense hydrogen \cite{Morales2013}. 

Figure~\ref{fig:phase-diagram} includes the predicted liquid-liquid phase transition (LLPT) line from CEIMC \cite{Pierleoni2016b}: the line intersects the melting line of the new isotropic phase at 1200K and 200GPa. Both the ML model and CEIMC predict dissociation as the LLPT line is crossed.  However the older CEIMC calculations, though the PES is more accurate (CEIMC does not make the assumptions that ML does), were done with simulations of 54 protons in a fixed orthorhombic volume: the boundary conditions and number of iterations did not allow the system to find the Fmmm-4 phase. A more complete comparison of the present results with those of CEIMC for the LLPT  will be presented elsewhere.
The present calculation corrects these biases and illustrates the power of combining QMC calculations with ML techniques.
However, additional research is required to make a definitive calculation of the hydrogen phase diagram. 
In particular, the QMC forces have been evaluated for systems with 96 protons and the ML descriptors are short ranged (see SM). Examination of the effect of the long-ranged interactions is needed. Of course, neither the ML description nor the QMC calculations are exact but both can be systematically improved. 
%It is also possible that the PIMD simulations did not find the equilibrium crystal structure, though having an even more stable structure would further increase the melting temperature. 
Finally, simulations were done without taking into account proton antisymmetry. This is expected to be a small effect at temperatures above 500K. Well-established techniques are available for its inclusion ~\cite{ceperley1996}.

%\ssec{Conclusions}

\ssec{Acknowledgment}
H. N. is thankful for the support provided by the NSF of China under Grants No. 12172112 and No. 11932005. The Flatiron Institute is a division of the Simons Foundation. D.M. C. and S. J. are supported by DOE DE-SC0020177. We thank Gábor Csányi, Linfeng Zhang, the QMC-HAMM team for valuable discussions and the QMCPACK team (J. Tiihonen and R. Clay III). We thank Kacper Kowalik and Matt Turk for helping organize our data in the yt Hub public database. Computations were done on Blue Waters Computer and the Illinois Campus Cluster, supported by the National Science Foundation (Awards No. OCI-0725070 and No. ACI-1238993), the state of Illinois, the University of Illinois at Urbana-Champaign, and its National Center for Supercomputing Applications. This research used resources of the Oak Ridge Leadership Computing Facility, which is a DOE Office of Science User Facility supported under Contract DE-AC05-00OR22725, and HPC resources from GENCI-IDRIS 2022-AD010912502R1.

\bibliographystyle{apsrev4-1}
\bibliography{ref}

\end{document}